\documentclass[prl,superscriptaddress,showpacs,twocolumn]{revtex4}
\usepackage{amsmath,graphicx,epsfig,color,amsfonts}
\usepackage[hypertex]{hyperref}

\newcommand{\ket}[1]{|#1\rangle}
\newcommand{\dd}{{\rm d}}
\newcommand{\GHZnt}{\ket{\mbox{\tiny GHZ}_n}}
\newcommand{\GHZn}{\ket{\rm GHZ}_n}
\renewcommand{\S}{\mathcal{S}}

\newcommand{\njp}{New J. Phys. }

\begin{document}

\title{Nonclassical Correlations from Randomly Chosen Local Measurements}

\author{Yeong-Cherng~Liang}
\email{ycliang@physics.usyd.edu.au} \affiliation{School of Physics, The
University of Sydney, Sydney, New South Wales 2006, Australia.}

\author{Nicholas Harrigan}
\affiliation{Department of Physics, Imperial College London, London SW7 2BW,
United Kingdom.}

\author{Stephen~D.~Bartlett}
\affiliation{School of Physics, The University of Sydney, Sydney, New South
Wales 2006, Australia.}

\author{Terry Rudolph}
\affiliation{Institute for Mathematical Sciences, Imperial College London,
London SW7 2BW, United Kingdom.}

\date{\today}
\pacs{03.65.Ud, 03.65.Ta, 03.67.Mn}

\begin{abstract}
We show that correlations inconsistent with any locally causal description can
be a generic feature of measurements on entangled quantum states. Specifically,
spatially-separated parties who perform local measurements on a
maximally-entangled state using {\em randomly chosen} measurement bases can,
with significant probability, generate nonclassical correlations that violate a
Bell inequality. For $n$ parties using a Greenberger-Horne-Zeilinger state,
this probability of violation rapidly tends to unity as the number of parties
increases. We also show that, even with both a randomly chosen two-qubit pure
state and randomly chosen measurement bases, a violation can be found about
10\% of the time. Amongst other applications, our work provides a feasible
alternative for the demonstration of Bell inequality violation without a shared
reference frame.
\end{abstract}

\maketitle

One of the most remarkable features of quantum mechanics is that distant
measurements can exhibit correlations that are inconsistent with any locally
causal description (LCD)~\cite{Bell}.  These quantum correlations are
signatures of entanglement~\cite{Horodecki:RMP:entanglement}, and serve as a
resource~\cite{Resource} for a range of classically-impossible information
processing tasks such as quantum key distribution~\cite{QKD},
teleportation~\cite{Teleportation} and reduced communication
complexity~\cite{Brukner:PRL}.

Not all quantum states are useful for quantum information processing tasks.
Even for entangled states, nonclassical correlations are not an immediate
consequence of the entanglement present in such systems  (see
Ref.~\cite{BIVvsEnt} and references therein).  They also depend crucially on
the choice of measurements to which they are subjected, and specifically how
these measurements are correlated between parties. For example, in the standard
scenarios to violate a Bell-Clauser-Horne-Shimony-Holt (Bell-CHSH)
inequality~\cite{Bell,CHSH} or more generally an $n$-party  inequality using a
Greenberger-Horne-Zeilinger (GHZ) state, a maximum violation is achieved by all
parties using specific measurements in the common $x-y$
plane~\cite{V.Scarani:JPA:2001}. In an experiment, this requires some
care~\cite{A.Aspect:0402001}; for two parties using polarization-encoded
single-photon pairs transmitted through long birefringent optical fibers, the
random rotation of the polarization of each photon is first compensated through
some method of alignment in order to set this frame~\cite{BRS:RMP}.  With
larger numbers of parties, greater violations can be achieved but more parties
require an increasing complexity of alignments.

Several approaches to circumvent this problem have been studied.  Encoded
entangled states that are invariant with respect to some collective unitary
operations~\cite{S.D.Bartlett:RFF,A.Cabello:RFFBIV} can be used, but generally
require much more complicated state preparation as well as joint measurements
on multiple spins.  Alternatively, an alignment of frames can be performed
through the coherent exchange of quantum
systems~\cite{RudolphGrover:PRL:2003,BRS:RMP} or by supplementing each system
with a small quantum reference frame of bounded size~\cite{F.Costa:0902.0935}.
Such solutions are resource-intensive given that this alignment consumes many
of these quantum resources that could otherwise be used to generate useful
correlations.

Here, we prove that such resource-intensive approaches are unnecessary, and
that nonclassical correlations occur with high probability even without
\emph{any} alignment of local measurements.  Specifically, we investigate the
detection of nonlocal correlations wherein each party uses randomly chosen
measurement bases. Ruling out LCD in this scenario cannot be achieved
deterministically for a general entangled quantum
state~\cite{A.Cabello:RFFBIV}. However, we demonstrate that the probability of
observing correlations inconsistent with LCD using randomly chosen measurement
bases can be remarkably high.  Although one might naively expect that
increasing the number of parties participating in the test (and the
corresponding complexity of alignments) would only complicate matters, we show
that the probability of demonstrating a violation rapidly approaches unity as
the number of parties increases.  Finally, we demonstrate the possibility of
two parties violating a Bell inequality using both randomly chosen measurements
and a randomly chosen pure state.

We consider a typical scenario of an $n$-party Bell-experiment using spin-$1/2$
particles, wherein a verifier prepares $N\gg1$ copies of the $n$-partite GHZ
state
\begin{equation}
    \label{eq:GHZ}
    \GHZn=\tfrac{1}{\sqrt{2}} \left(\ket{0}_1\ket{0}_2\cdots\ket{0}_n
    +\ket{1}_1\ket{1}_2\cdots\ket{1}_n \right)\,,
\end{equation}
and distributes them to $n$ parties.  Each particle is subjected to a local
measurements with two possible outcomes, $\pm1$.  The measurement for each
particle is chosen randomly, with equal probability, from a set of two
measurement bases, defined for the $k$th party by a pair of spatial directions
$\Omega^{[k]}_{s_k}(\theta,\phi)$, with $s_k=1,2$ labeling the bases.  For
spins measured via a Stern-Gerlach device, these directions correspond to the
orientation of the magnetic field in the measurement apparatus.  However, an
equivalent picture in terms of spatial directions can be used for any quantum
systems described by a two-dimensional Hilbert space (eg., the polarization
encoding of single photons) via the Bloch sphere. After the completion of all
measurements, the $n$ parties return to the verifier a list containing only the
label $s_k$ of which measurement they performed for each particle and the value
$\pm 1$ of the outcome they obtained for that measurement.

A relevant Bell inequality for this scenario is the $n$-partite
Mermin-Ardehali-Belinski\v{\i}-Klyshko (MABK)
inequality~\cite{MerminIneq,ZB:Complete:npartiteBI, WW:Complete:npartiteBI},
\begin{equation}
    \S^{(n)}_{\mbox{\tiny MABK}}
    =\Bigl|\sum_{s_1,\ldots,s_n=1}^2\beta(s_1,\ldots,s_n)
    E(s_1,\ldots,s_n)\Bigr|\le2^{n-\frac{1}{2}},
    \label{Ineq:MABK}
\end{equation}
where
\begin{equation}\label{Eq:MABK:Coeff}
    \beta(s_1,\ldots,s_n)=\!\!\!\!\!\!\!\sum_{k_1,\ldots,k_n=\pm1}
    \!\!\!\!\cos\Bigl[\frac{\pi}{4}\bigl(n+1-\sum_{l=1}^n
    k_l\bigr)\Bigr]\prod_{j=1}^n k_j^{s_j-1}.
\end{equation}
Here, the $n$-partite correlation function $E(s_1,\ldots,s_n)$ is  defined to
be the expectation value of the product of measurement outcomes ($\pm1$) given
that the $k$th party measured their spin in the basis ${s_k}$.  When $n=2$, the
MABK inequality is equivalent to the familiar Bell-CHSH inequality
$\S_{\mbox{\tiny CHSH}} =|E(1,1)+E(1,2)+E(2,1)-E(2,2)|\le 2$~\cite{Bell,CHSH}.

In a standard Bell experiment, the maximal violations for a given Bell
inequality are achieved by choosing the local orientations $\Omega^{[k]}_{1}$
and $\Omega^{[k]}_{2}$ to be perpendicular~\cite{V.Scarani:JPA:2001}, i.e.,
$\Omega^{[k]}_{1}\cdot\Omega^{[k]}_{2}=0$, and by fixing the relative
orientations of these measurements with the other experimentalists (and the
verifier who prepared the systems).  The latter necessitates the use of a
shared reference frame; in what follows, we will see what can be achieved
without a shared reference frame.

We will first consider the completely unconstrained case, wherein each party
$k$ chooses both members of their pair of directions $\Omega^{[k]}_{s_k}$ for
$s_k=1,2$ independently and uniformly from the set of all possible directions.
We refer to this case as \emph{random isotropic measurements} (RIM).  In such a
scenario, we  determine the probability $p^{\GHZnt}_{\mbox{\tiny MABK}}$ that
the measurement statistics received by the verifier will give rise to a
violation of the MABK inequality of Eq.~(\ref{Ineq:MABK}), given by
\begin{equation}\label{Eq:Prob:Dfn:Mermin}
    p^{\GHZnt}_{\mbox{\tiny MABK}}
    =\frac{1}{(4\pi)^{2n}}\!\!\int\prod_{k=1\ldots n\atop  s_k=1,2}\dd\Omega^{[k]}_{s_k}
    f^{\mbox{\tiny MABK}}_n (\{\Omega^{[k]}_{s_k}\}_{k=1}^n )\,,
\end{equation}
where $\dd\Omega^{[k]}_{s_k}/4\pi$ is the Haar measure associated with
measurement direction $\Omega^{[k]}_{s_k}$ and $f^{\mbox{\tiny MABK}}_n
(\{\Omega^{[k]}_{s_k}\}_{k=1}^n )$ is a function that returns 1 if orientations
$\{\Omega^{[k]}_{s_k}\}_{k=1}^n$ give rise to a correlation (i.e., measurement
statistics), that violates the MABK inequality and 0 otherwise. These integrals
simply pick out those measurement directions that will give a violation,
normalizing them against the total volume of possible measurement directions.

For the detection of nonclassical correlation, we can allow the verifier to
make use of the freedom in relabeling all measurement settings and/or outcomes
and/or parties~\cite{WW:Complete:npartiteBI, EquivalentBI}. To appreciate the
importance of this fact, consider as an example the following choice of local
measurement for the $n=2$ case: $\{\sigma_z,\sigma_x\}$ for Alice and
$\{\frac{1}{\sqrt{2}}\left(\sigma_x+\sigma_z\right),
\frac{1}{\sqrt{2}}\left(\sigma_x-\sigma_z\right)\}$ for Bob. Suppose that Alice
has labeled her measurements such that $\sigma_z \leftrightarrow s_1{=}1$ and
$\sigma_x \leftrightarrow s_1{=}2$, while Bob has labeled his measurements such
that $\frac{1}{\sqrt{2}}\left(\sigma_x+\sigma_z\right) \leftrightarrow s_2{=}1$
and $\frac{1}{\sqrt{2}}\left(\sigma_x-\sigma_z\right) \leftrightarrow s_2{=}2$,
then the verifier will find that \emph{as it is}, $\S_{\mbox{\tiny CHSH}}$
vanishes, and hence does not lead to a Bell-CHSH inequality violation. However,
if the verifier is also given the freedom to relabel the measurement settings,
they could use, instead, the following labeling for Alice's measurements:
$\sigma_x \leftrightarrow s_1{=}1$ and $\sigma_z \leftrightarrow s_1{=}2$. As
opposed to the original labeling, this new labeling unveils measurement
statistics that violate the Bell-CHSH inequality maximally.  We say that two
Bell inequalities are \emph{equivalent} if they can be obtained from each other
by relabeling the measurement settings and/or outcomes and/or
parties~\cite{WW:Complete:npartiteBI,EquivalentBI}.  In calculating the
probability of observing a violation, we clearly want to include all such
equivalent inequalities.

We numerically calculate $p^{\GHZnt}_{\mbox{\tiny MABK}}$, using the relabeling
strategy to test a given set of experimental data against all $2^{n+1}$
equivalent MABK inequalities~\cite{fn:AllMABK}; see
Fig.~\ref{Fig:CombinedEverything}. The special case of $n=2$ can be solved
analytically (see supplementary materials attached below for details) and is
found to be $2(\pi-3)\approx 28.3\%$ (see also
Ref.~\cite{P.Lougovski:PRA:034302} for contrasting results). For $n>2$, the
probability is less than this amount but increases monotonically for $n>3$ up
to the limit of our analysis, $n=15$. Note, however, it appears that this
probability is asymptotically approaching a value that is less than unity.

\begin{figure}[h!]
\scalebox{0.5}{\includegraphics{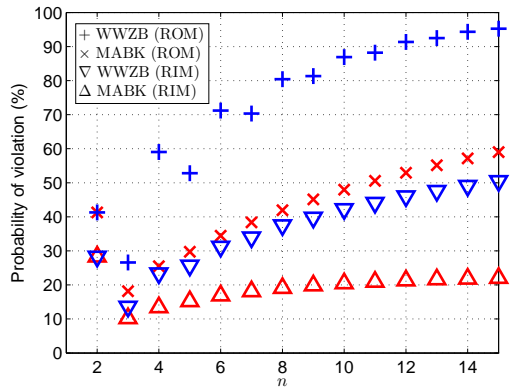}}
\caption{\label{Fig:CombinedEverything}
(Color online) Probability of finding a nonclassical $n$-partite GHZ
correlation. The two sets of data at the bottom and top are, respectively, for
correlation generated from {\em random isotropic measurements} (RIM) and {\em
random orthogonal measurements} (ROM). The markers are for the scenarios when
the verifier has access to (1) the $2^{n+1}$ MABK inequalities ($\Delta$ for
RIM, $\times$ for ROM), and (2) the $2^{2^n}$ WWZB inequalities ($\nabla$ for
RIM, $+$ for ROM).}
\end{figure}

The case of $n=2$ appears somewhat anomalous.  The reason is that, for $n=2$,
the equivalence class of Bell-CHSH inequalities obtained by relabeling
$\S^{(2)}_{\mbox{\tiny MABK}}$ (or $\S_{\mbox{\tiny CHSH}}$) is both necessary
and sufficient for determining if a given set of experimental statistics
generated by performing two binary-outcome measurements per site admits a
LCD~\cite{A.Fine:PRL:1982}. In other words, in the space of measurement
statistics generated by performing such measurements, all nonclassical
correlations can be detected by this equivalence class of inequalities.

For $n>2$, there are \emph{(i)} Bell inequalities with $n$-partite correlation
functions that are \emph{not} of the MABK form  (\ref{Ineq:MABK}) and
\emph{(ii)} Bell inequalities involving less-than-$n$-partite correlation
functions.  To determine if a given correlation for $n>2$ is incompatible with
\emph{any} LCD, we need to test the measurement statistics against the complete
set of Bell inequalities that are relevant to the particular experiment. For a
small number of parties, this can be achieved efficiently using {\em linear
programming} (see~\cite{D.Gosal:PRA:042106} and the supplementary materials
attached below) or by checking the correlation directly against the complete
set of relevant Bell inequalities. Our results are summarized in
Table~\ref{tbl:ProbViolation}.

\begin{table}[t]
\begin{ruledtabular}
\begin{tabular}{c||c|c|c|c|c}
$n$ & 2& 3 & 4  &5 &6 \\ \hline
RIM & 28.3185\% & 74.6899\% & 94.2380\% & 99.5926\% & $99.97\%$   \\
$\tfrac{\text{Run}}{10^5}$ &    - & 5000  & 500 & 25 & 0.8\\ \hline
ROM & 41.2982\% & 96.2073\% & 99.9757\% & $99.9999\%$ & $100.00\%$ \\
$\tfrac{\text{Run}}{10^5}$ & 5000 & 5000  & 500 & 25 & 0.8\\
\end{tabular}
\end{ruledtabular}
\caption{\label{tbl:ProbViolation} Probability of finding a nonclassical
correlation from the $n$-partite GHZ state for the scenario where each party is
allowed to perform binary projective measurements in two randomly chosen
measurement bases. The number of data points (Run) used to compute the
probabilities for RIM (ROM) is included in the third (fifth) row of the table.}
\end{table}

These results clearly indicate that the probability of a violation rapidly
approaches unity with increasing $n$.  The input to the linear program,
however, scales exponentially with $n$ (see supplementary materials attached
below), thus making it intractable to determine the probability of violation
reliably for larger values of $n$. To obtain some insight into the behavior, we
can restrict our attention to the extensive set of $2^{2^n}$ $n$-partite
correlation inequalities discovered independently by Werner \&
Wolf~\cite{WW:Complete:npartiteBI} and \.Zukowski \&
Brukner~\cite{ZB:Complete:npartiteBI} (WWZB). While violation of these
inequalities is still sufficient to rule out LCD, the converse is generally not
true. Thus, using this set of inequalities yields a lower bound on the
probability that a given correlation is nonclassical.

To test if a given correlation satisfies all of the WWZB inequalities is
equivalent to testing if it satisfies the following nonlinear
inequality~\cite{WW:Complete:npartiteBI,ZB:Complete:npartiteBI}
\begin{equation}\label{Ineq:nonlinear}
    \sum_{k_1,\ldots,k_n=\pm1}\Bigl|\sum_{s_1,\ldots,s_n=1}^2
    \prod_{j=1}^n k_j^{s_j-1}E(s_1,\ldots,s_n)\Bigr|\le2^{n}.
\end{equation}
We have numerically computed the probability of finding a randomly generated
correlation from RIM to violate inequality~\eqref{Ineq:nonlinear} for $n\le
15$. As can be seen in Fig.~\ref{Fig:CombinedEverything}, this probability
generally increases with $n$ and is above 50\% for $n=15$.  However, it is
inconclusive whether this probability asymptotes to unity or not for large $n$.

Finally, we consider another mechanism by which parties who do not share a
reference frame can increase the probability of observing a violation.  So far
we have considered each of the two local measurement bases for each party to be
chosen independently and isotropically. However, because the optimal violations
are typically found when the pair of local measurement directions are
orthogonal~\cite{V.Scarani:JPA:2001}, we can consider a random selection of
measurement bases under the constraint that each local pair are orthogonal. We
have repeated the above calculations with this constraint (i.e., that
$\Omega^{[k]}_1\cdot \Omega^{[k]}_2=0$); we refer to this distribution as
\emph{random orthogonal measurements} (ROM).  The results reveal a significant
increase in the probability of a violation, as shown in
Fig.~\ref{Fig:CombinedEverything} and Table~\ref{tbl:ProbViolation}. For larger
values of $n$, our results for the WWZB inequalities also strongly suggest that
the probability asymptotes to unity for large $n$.

Our results so far have all made use of a fixed maximally-entangled pure state
(\ref{eq:GHZ}) (although clearly the local bases in which this state is defined
are irrelevant for our result).  We can also consider how the probability of
observing a violation of LCD varies with the degree of entanglement of the
distributed quantum state. For $n=2$, a pure quantum state can always be
written as $\ket{\Psi}=\cos\theta\ket{0}_1\ket{0}_2
+\sin\theta\ket{1}_1\ket{1}_2$ for some local bases.  Numerically, we can
compute the probability of violation as a function of entanglement (for
details, see supplementary materials attached below).  With this result, we can
determine the probability of observing a violation of LCD using both randomly
chosen measurement bases and a randomly chosen state. We sample two-qubit pure
states randomly and uniformly from $\mathbb{C}^2\otimes\mathbb{C}^2$ according
to the SU(4) Haar measure. Given that our results are independent of local
choice of bases, we can reproduce this distribution by sampling Schmidt
coefficients as for single-qubit mixed states from the uniform Bloch
ball~\cite{I.Bengtsson:Book}. The relevant measure is the Hilbert-Schmidt
measure on a single qubit, i.e., $P_{\mbox{\tiny HS}}(r)=24r^2$, where
$r=\frac{\cos2\theta}{2}$ is the length of the Bloch vector (pp.~354,
Ref.~\cite{I.Bengtsson:Book}).  The probability of violation for a randomly
chosen pure two-qubit state can then be determined by numerical integration to
be, respectively, about 5.3\% for RIM and 10.1\% for ROM.

{\em Conclusion and other directions.}-- We introduced the idea of performing a
Bell experiment using randomly-chosen measurement bases, with the aim of
demonstrating nonclassical correlations in the absence of a shared reference
frame. We applied this idea to the $n$-partite GHZ state and showed that the
probability of finding a nonclassical correlation rapidly tends to unity as the
number of parties increases. Naively, one might have expected that this chance
would diminish with increasing $n$, due to the increasing chance of
misalignment of the experimenters' measurement bases (or, alternatively, by
considering the fragility of the GHZ state to dephasing). Our results clearly
show otherwise.  We have thus showed that without a shared reference frame, a
Bell inequality violation can still be demonstrated reliably without resorting
to complicated state preparation or consumption of expensive quantum resources.
This, of course, significantly reduces the technical requirements for
experimentally violating a Bell inequality, performing quantum key distribution
based on such violations, or establishing large-scale quantum networks.

Our work also represents the first systematic study of the set of correlation
derivable from a quantum resource such as the $n$-partite GHZ state.
Specifically, our results indicate that the correlations obtained by performing
projective measurements on this quantum resource are almost ubiquitously
nonclassical (this can be rigorously quantified by considering a volume measure
in the space of GHZ-attainable-correlations induced by RIM and ROM). Given that
measurements and the resulting correlations play an important role in many
quantum information processing tasks, the tools and ideas that we have
introduced here, suitably generalized, may shed some light on the origin of the
power of other quantum resources such as the highly entangled cluster state. In
particular, they may add new insights to the computational power of
correlations recently discussed in Ref.~\cite{J.Anders:PRL:050502}.

The results presented here motivate several additional research directions.
First, one may attempt to determine a new intuition for our result of
increasing probability of violation with the number of parties $n$, by relating
this probability to the average magnitude of the MABK violation, which also
increases with $n$.  Second, for many quantum information processing tasks,
violation of the inequalities considered in the present work may not be
sufficient, as other types of multipartite entanglement (perhaps different from
that of the GHZ state) may be required that are not quantified by these
inequalities.  To this end, it will be interesting to investigate the
efficiency of random measurements against inequalities that detect other, more
general types of multipartite entanglement~\cite{FullEntanglement}.  Finally,
for real-world applications of these results, a critical issue is to determine
how robust they are against decoherence.

\begin{acknowledgments}
YCL acknowledges fruitful discussion with Nicolas Brunner, Tam\'as V\'ertesi,
Stefano Pironio, Mafalda Almeida, Valerio Scarani, Eric Cavalcanti and Daniel
Cavalcanti. SDB acknowledges support from the ARC. TR and NH acknowledge the
support of the EPSRC.
\end{acknowledgments}


\section{Supplementary Materials}

\subsection{MABK-violation without classical post-processing}

\begin{figure}[h!]
\scalebox{0.50}{\includegraphics{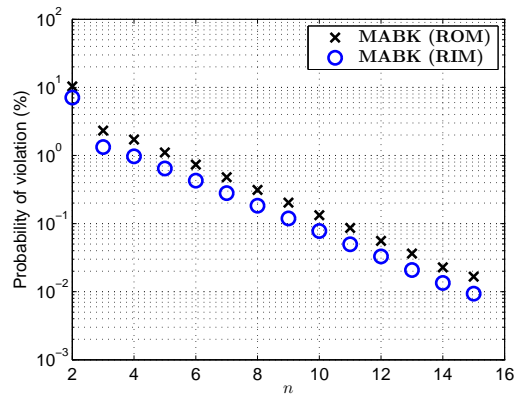}}
\caption{\label{Fig:ProbMABK}
(Color online) Probability of finding a randomly generation correlation from
$\GHZn$ to violate a single MABK inequality.}
\end{figure}

In the main text, we have presented the probability of observing a violation by
considering all equivalent Mermin-Ardehali-Belinski\v{\i}-Klyshko (MABK)
inequalities. If the classical post-processing to optimize over the measurement
labelings is \emph{not} carried out (i.e., we calculate the probability to
violate a particular inequality rather than an equivalent set), then we find
$p^{\GHZnt}_{\mbox{\tiny MABK}}$ decreases steadily with $n$; see
Fig.~\ref{Fig:ProbMABK}. However, in this case, among those choices of bases
that give rise to an MABK-violation, we find that their average violation
increases with $n$ when normalized by the classical threshold
$2^{n-\frac{1}{2}}$ but decreases when normalized by the maximal possible
quantum violation
$2^{\frac{3n}{2}-1}$~\cite{MerminIneq,WW:Complete:npartiteBI}. See also
Ref.~\cite{I.Pitowsky:PRA:022103} in this regard.

\subsection{Analytic derivation of probability of violation for $n=2$}

For the specific case of $n=2$, the probability of violating the (2-party) MABK
inequality, i.e., the Bell-CHSH inequality~\cite{Bell,CHSH}
\begin{equation}\label{Ineq:CHSH}
    \S_{\mbox{\tiny CHSH}} =|E(1,1)+E(1,2)+E(2,1)-E(2,2)|\le 2,
\end{equation}
can be computed analytically. To this end, it is expedient to consider the
two-qubit singlet state
\begin{equation}
    \ket{\Psi^-}=\tfrac{1}{\sqrt{2}} \left(\ket{0}_1\ket{1}_2-\ket{1}_1\ket{0}_2 \right),
\end{equation}
instead of the 2-party GHZ state:
\begin{equation}
    \ket{{\rm GHZ}}_2=\tfrac{1}{\sqrt{2}}\left(\ket{0}_1\ket{0}_2+\ket{1}_1\ket{1}_2\right).
\end{equation}
Clearly, these states are local unitarily equivalent, and hence their
probability of violating the Bell-CHSH inequality via random isotropic
measurements (ROM) is identical. However, $\ket{\Psi^-}$ is also rotationally
invariant (i.e., $U\otimes U \ket{\Psi^-}=\ket{\Psi^-}$ for arbitrary qubit
unitary transformation $U$), which allows us to simplify our analysis
considerably.

Let us begin by writing the measurement directions for Alice and Bob as
$\vec{a}_{s_1}$ and $\vec{b}_{s_2}$. From some simple calculation, one can show
that the correlation function, i.e., the average of the product of measurement
outcomes for Alice measuring $\vec{a}_{s_1}\cdot\vec{\sigma}$ and Bob measuring
$\vec{b}_{s_s}\cdot\vec{\sigma}$ is
\begin{equation}\label{Eq:Qubit:Correlation}
    E\left(\vec{a}_{s_1},\vec{b}_{s_2}\right)=-\vec{a}_{s_1}\cdot\vec{b}_{s_2},
\end{equation}
where $\vec{\sigma}=(\sigma_x, \sigma_y, \sigma_z)$ is the vector of Pauli
matrices. Using this in Eq.~\eqref{Ineq:CHSH}, we are thus interested to know
how often the inequality
\begin{equation}\label{bell_random_directions}
    \left|\vec{a}_1\cdot\vec{b}_1+\vec{a}_1\cdot\vec{b}_2+\vec{a}_2\cdot\vec{b}_1-\vec{a}_2\cdot\vec{b}_2\right|>2.
\end{equation}
holds when the unit vectors $\vec{a}_1$, $\vec{a}_2$, $\vec{b}_1$ and
$\vec{b}_2$ are chosen independently, randomly and isotropically. We can
simplify this by defining the two orthogonal vectors,
$\vec{r}_{\perp}=\vec{b}_1-\vec{b}_2$ and $\vec{r}=\vec{b}_1+\vec{b}_2$, in
terms of which inequality~\eqref{bell_random_directions} holds if and only if
\begin{equation}\label{bell_ineq_ar}
    \left|\vec{a}_1\cdot\vec{r}+\vec{a}_2\cdot\vec{r}_{\perp}\right|>2.
\end{equation}
Note that as well as being orthogonal, the vectors $\vec{r}_{\perp}$ and
$\vec{r}$ have lengths satisfying,
\begin{align}
    \left|\vec{r}\right|=\sqrt{2(1+\vec{b}_1\cdot\vec{b}_2)}
    =\sqrt{2(1+x)},\nonumber\\
    \left|\vec{r}_{\perp}\right|=\sqrt{2(1-\vec{b}_1\cdot\vec{b}_2)}
    =\sqrt{2(1-x)},\label{Eq:r_rp:magnitude}
\end{align}
where $x=\vec{b}_1\cdot\vec{b}_2$.

By rewriting $\vec{r}$ in terms of its unit vector $\hat{r}$ and magnitude
$|\vec{r}|$, i.e., $\vec{r}=\hat{r}|\vec{r}|=\hat{r}\sqrt{2(1+x)}$, and
similarly for $\vec{r}_\perp$, inequality~\eqref{bell_ineq_ar} becomes
\begin{equation}\label{Ineq:arx}
    \left|\vec{a}_1\cdot\hat{r}\sqrt{1+x}+\vec{a}_2\cdot\hat{r}_{\perp}\sqrt{1-x}
    \right|>\sqrt{2}.
\end{equation}
Now the rotational invariance of dot products ensures that in determining the
fraction of measurement directions that violate inequality~\eqref{Ineq:arx},
the actual direction of $\hat{r}$ and $\hat{r}_\perp$ is irrelevant. With some
thought, one can see that to determine this fraction, and hence the probability
of violating the Bell-CHSH inequality~\eqref{Ineq:CHSH} via random isotropic
measurements, it suffices to sample $x$ and the dot products
$\vec{a}_1\cdot\hat{r}$, $\vec{a}_2\cdot\hat{r}_{\perp}$ uniformly from the
interval $[-1,1]$. Writing $\beta=\vec{a}_1\cdot\hat{r}$ and
$\alpha=\vec{a}_2\cdot\hat{r}_{\perp}$, we therefore wish to find the fraction
of values of $\alpha,\beta,x\in[-1,1]$ satisfying,
\begin{equation}\label{random_bell_alphabetax}
    \left|\alpha\sqrt{1-x}+\beta\sqrt{1+x}\right|>\sqrt{2}.
\end{equation}

This problem has a useful geometrical interpretation. The set of all points
$\alpha,\beta,x\in[-1,1]$ define a cube, $\mathfrak{C}$, which encloses a
volume $2^3$ and has faces located at $\alpha,\beta,x=\pm{1}$. To find the
fraction of Alice and Bob's measurement directions that would violate the
Bell-CHSH inequality we need to find what fraction of the volume of
$\mathfrak{C}$ contains points $(\alpha,\beta,x)$ satisfying the constraint of
Eq.~(\ref{random_bell_alphabetax}).

\begin{figure*}
\begin{tabular}{ccccc}
\includegraphics[scale=0.15]{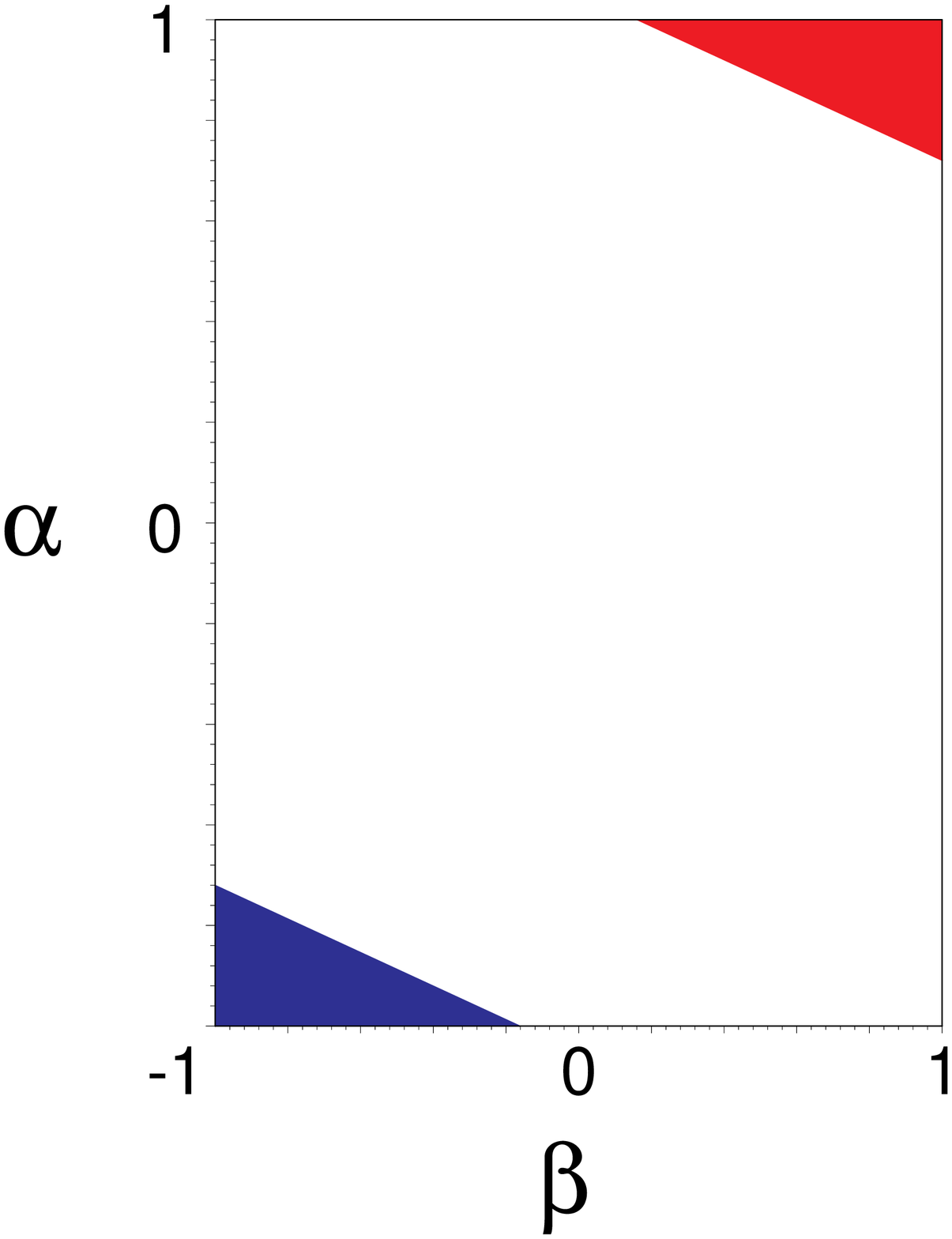}&
\includegraphics[scale=0.15]{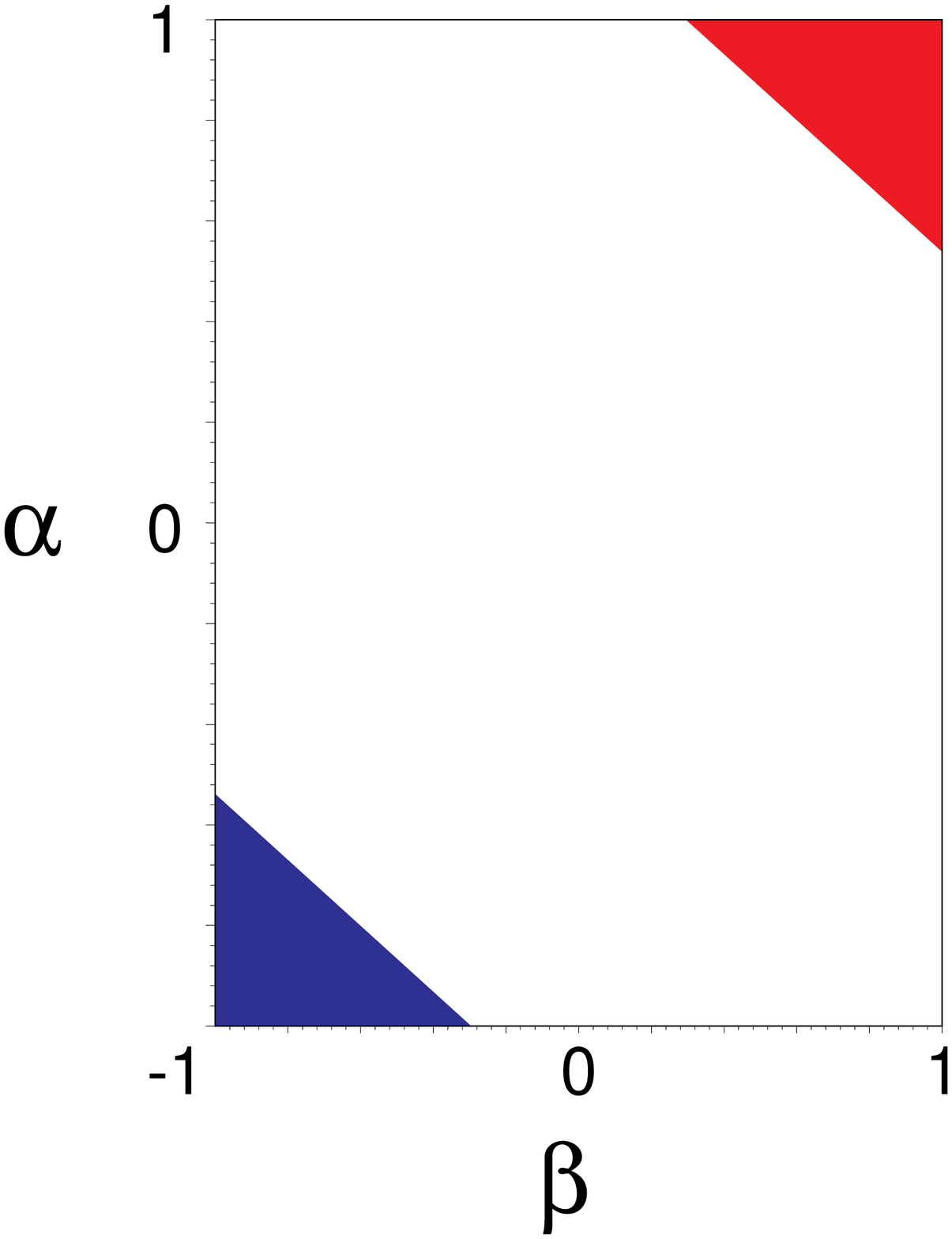}&
\includegraphics[scale=0.15]{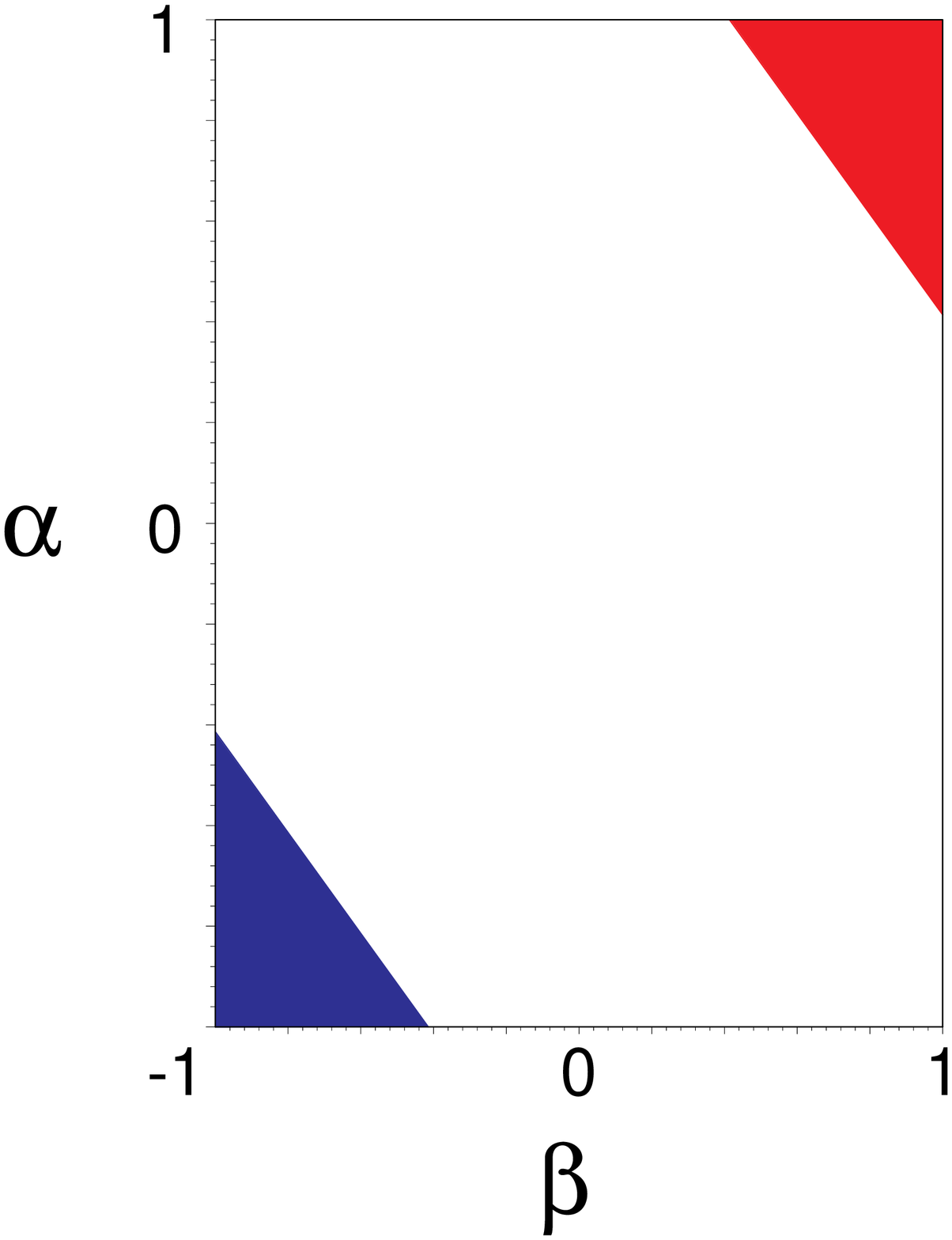}&
\includegraphics[scale=0.15]{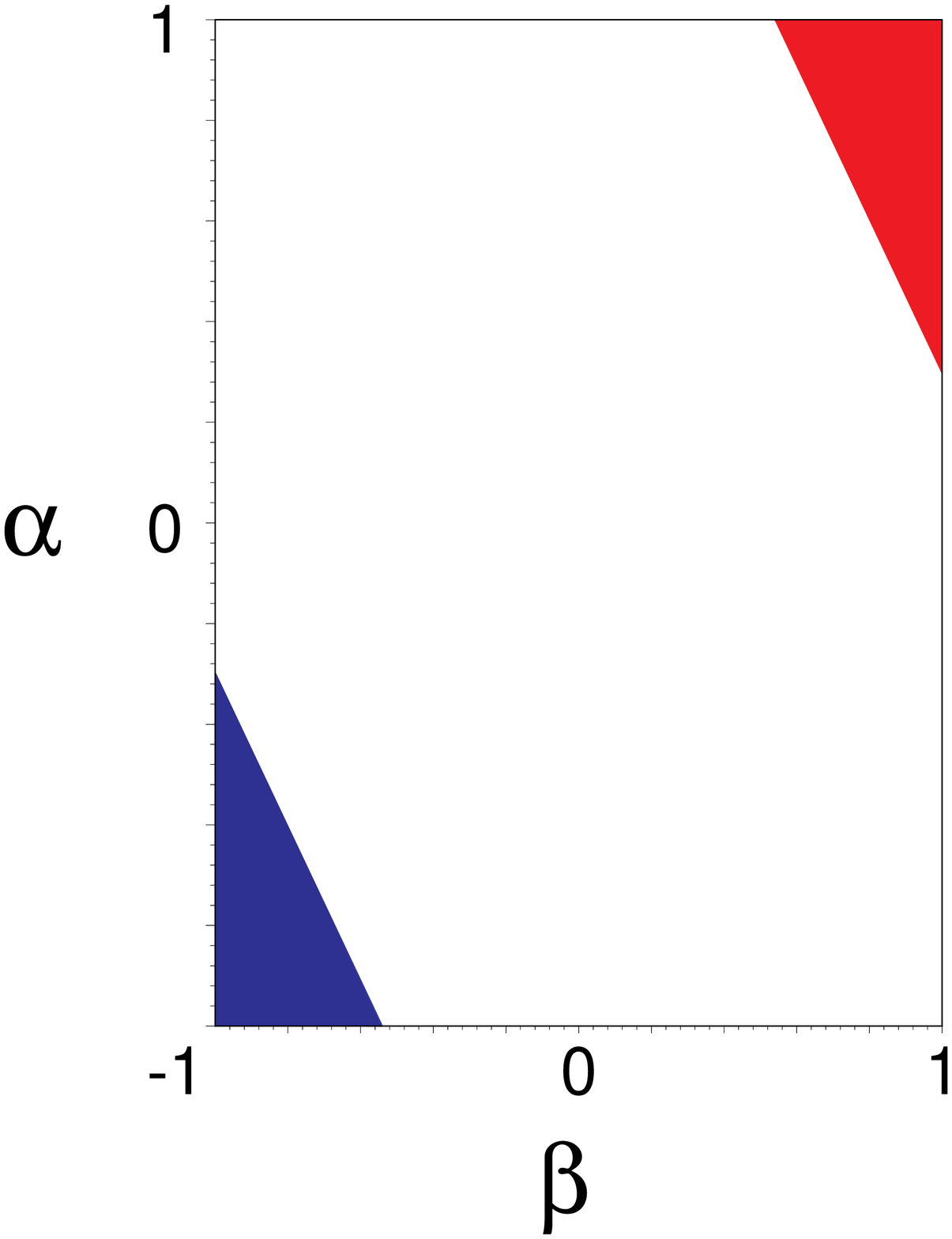}&
\includegraphics[scale=0.15]{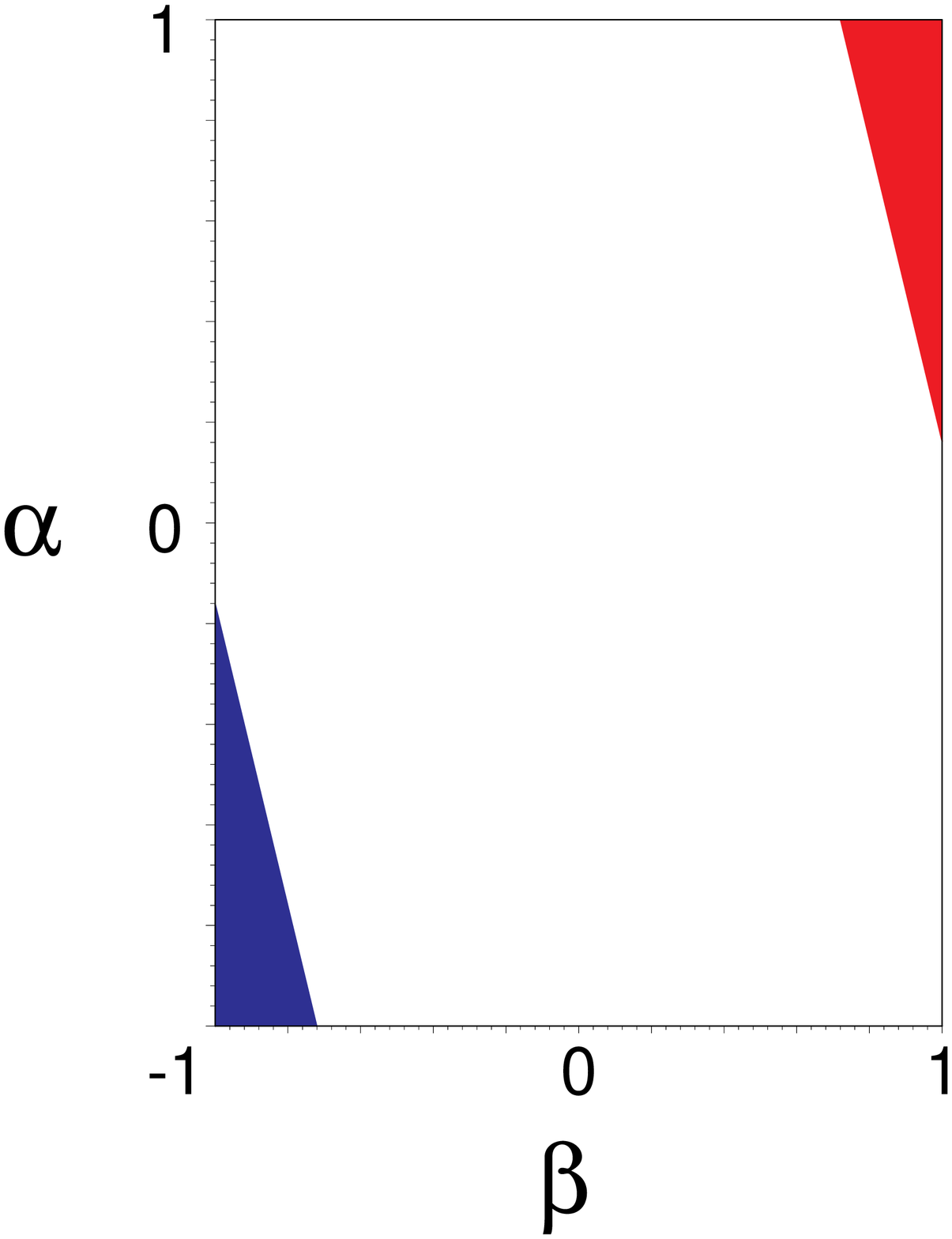}
\end{tabular}
\caption{(Color online) Plots showing the triangular regions containing values
of $\alpha$ and $\beta$ which satisfy inequality~\eqref{random_bell_alphabetax}
(i.e., violating the Bell-CHSH inequality). From left to right, the plots are
for values of $x=-0.8,-0.4,0,0.4,0.8$.\label{FIG:cube_cross_sections} }
\end{figure*}


By separately considering the cases where $\alpha\sqrt{1-x}+\beta\sqrt{1+x}>0$
and $\alpha\sqrt{1-x}+\beta\sqrt{1+x}<0$, the regions of $\mathfrak{C}$
containing points satisfying Eq.~(\ref{random_bell_alphabetax}) are given by,
\begin{equation}\label{triangle_1}
    \alpha>\frac{\sqrt{2}-\beta\sqrt{1+x}}{\sqrt{1-x}}
\end{equation}
and,
\begin{equation}\label{triangle_2}
    \alpha<-\frac{\sqrt{2}+\beta\sqrt{1+x}}{\sqrt{1-x}}.
\end{equation}

If we take a cross-section through $\mathfrak{C}$ corresponding to a particular
fixed value of $x$, then these two expressions define triangular regions at the
corners of the cube, as is shown in Fig.~\ref{FIG:cube_cross_sections}. From
some straightforward computation, it can be shown that the total area of the
two triangles for a given $x$ is
\begin{equation}
\frac{\left(\sqrt{1+x}+\sqrt{1-x}-\sqrt{2}\right)^2}{\sqrt{1-x^2}}.
\end{equation}
Integrating this expression from $x=-1$ to 1, and dividing the result by $2^3$
(the total volume of $\mathfrak{C}$), we therefore find that the fraction of
Alice and Bob's measurement directions that will violate the Bell-CHSH
inequality is:
\begin{equation}
    \frac{\pi-3}{2}\simeq{7.0796\%}.
\end{equation}

Notice that the above expression is the probability of violating
inequality~\eqref{Ineq:CHSH} without taking into account of any possible
relabeling of measurement settings and/or outcomes. To take this into account,
note that the four equivalent Bell-CHSH inequalities are as follows:
\begin{align}\label{Ineq:CHSH:Equivalent}
    |E(1,1)+E(1,2)-E(2,1)-E(2,2)|\le 2,\\
    |E(1,1)+E(1,2)-E(2,1)+E(2,2)|\le 2,\\
    |E(1,1)-E(1,2)+E(2,1)+E(2,2)|\le 2,\\
    |-E(1,1)+E(1,2)+E(2,1)+E(2,2)|\le 2.
\end{align}
It is not difficult to see that for any given measurement directions, {\em at
most} one of these inequalities can be violated. To see this, suppose otherwise
and one will find that an inequality like the following
\begin{equation}
    E(1,1)+E(1,2)>2,
\end{equation}
has to hold true. However, this is logically impossible as the absolute value
of each of the correlation function $|E(s_1,s_2)|$ is upper bounded by 1.

Moreover, by symmetry, the probability of finding measurement directions that
violate any of these equivalent inequalities is identical. Hence, the
probability of finding measurement directions that violate any of the Bell-CHSH
inequality, with the help of relabeling of measurement settings and/or outcomes
is:
\begin{equation}
    4\times\frac{\pi-3}{2}\simeq{28.3185\%}.
\end{equation}

\subsection{Identifying nonclassical correlation via linear programming}

Linear programming is a convex optimization problem~\cite{S.Boyd:Book:2004}
which can be efficiently solved on a computer. For this kind of optimization
problem, both the objective function and the constraints are linear in the
variable $\mathbf{x}\in\mathbb{R}^d$. In its {\em standard form}, a linear
program reads as~\cite{S.Boyd:Book:2004}:
\begin{subequations}
\begin{align}
    &\,\text{~minimize}\qquad \mathbf{c}^{\mbox{\tiny T }}\mathbf{x},\label{Eq:Obj}\\
    &\text{~subject to}\quad A\mathbf{x}=\mathbf{b},\label{Eq:Const:Eq}\\
    &\qquad\qquad\qquad\,\,\, \mathbf{x} \succeq\mathbf{0}\label{Eq:Const:Ineq},
\end{align}
\end{subequations}
where $\mathbf{c}\in\mathbb{R}^{d}$, $\mathbf{b}\in\mathbb{R}^{d'}$, $A$ is a
$d'\times d$ matrix, $\succeq$ represents a component-wise inequality and
$\mathbf{0}$ is a $d\times1$ null vector.

For the experimental scenarios described in the main text, the $k$-th party is
allowed to perform measurements in two different measurement directions
$\Omega^{[k]}_{s_k}$, with $s_k=1,2$. Let us denote by $o^{[k]}_{s_k}$ the
measurement outcome corresponding to the measurement direction
$\Omega^{[k]}_{s_k}$. Now, recall that a given correlation admits a locally
causal/ realistic description if and only if all local measurement outcomes
$o^{[k]}_{s_k}$ are (probabilistically) determined in advance solely by the
choice of the local setting $s_k$. Equivalently, a locally causal correlation
is one where the experimental statistics admit a joint probability distribution
\begin{equation*}
    p_{1\ldots n}(o^{[1]}_1, o^{[1]}_2, \ldots, o^{[k]}_1,o^{[k]}_2,\ldots, o^{[n]}_1,
    o^{[n]}_2)
\end{equation*}
for all the $2n$ variables $o^{[k]}_{s_k}$, and which recovers all the observed
experimental statistics as marginals~\cite{fn:ExplicitBipartite}. For example,
the joint probability $p_{1n}$ of observing $o^{[1]}_1=1$ and $o^{[n]}_2=-1$,
must be reproducible via:
\begin{align*}
    &p_{1n}(o^{[1]}_1=1,o^{[n]}_2=-1)\\
    =&\sum_{k=2}^{n-1}\sum_{o^{[1]}_2,o^{[n]}_1,o^{[k]}_{s_k}=\pm1}
    p_{1\ldots n}(o^{[1]}_1, o^{[1]}_2, \ldots, o^{[k]}_1,o^{[k]}_2,\ldots, o^{[n]}_1,
    o^{[n]}_2);
\end{align*}
likewise for all the correlation functions, such as
\begin{align*}
    &E(s_1,s_n)=\sum_{k=1,n}\sum_{o^{[k]}_{s_k}=\pm1}
    \prod_{j=1,n} o^{[j]}_{s_j} \,p_{1n}(o^{[1]}_{s_1},o^{[n]}_{s_n}),
\end{align*}
which must also be reproducible from appropriate linear combination of the
entries of the joint probability distribution $p_{1\ldots n}$ (see
Ref.~\cite{D.Gosal:PRA:042106} for a more elaborate exposition of these
constraints).

Clearly, these constraints can be put in the form of Eq.~\eqref{Eq:Const:Eq},
with $\mathbf{x}\in\mathbb{R}^{2^{2n}}$ being the entries of $p_{1\ldots n}$
sought for, with all (but one) entries of $\mathbf{b}\in\mathbb{R}^{3^n}$ being
the measured correlation functions $E(s_1,\ldots, s_k)$, and with entries of
$A$ giving the coefficients required to reproduce each $E(s_1,\ldots, s_k)$
from the entries of $p_{1\ldots n}$. On top of the $3^{n}-1$ correlation
functions, $\mathbf{b}$ must also contain an additional entry that is set to 1;
the same goes for the corresponding row of entries of $A$, which must all be
set to 1. This last requirement ensures that the joint probability distribution
$p_{1\ldots n}$ is normalized while the non-negativity of the probabilities is
taken care of by Eq.~\eqref{Eq:Const:Ineq}.

The problem of determining if a given correlation admits a locally causal
description can then be cast as the following linear programming {\em
feasibility problem}:
\begin{align*}
    &\,\text{~minimize}\qquad\quad\, 0,\\
    &\text{~subject to}\quad A\mathbf{x}=\mathbf{b},\\
    &\qquad\qquad\qquad\,\,\, \mathbf{x} \succeq\mathbf{0}.
\end{align*}
For an alternative formulation of the problem as a linear programming {\em
optimization problem}, see Ref.~\cite{M.B.Elliott:0905.2950}.

\begin{figure}[h!]
\scalebox{0.50}{\includegraphics{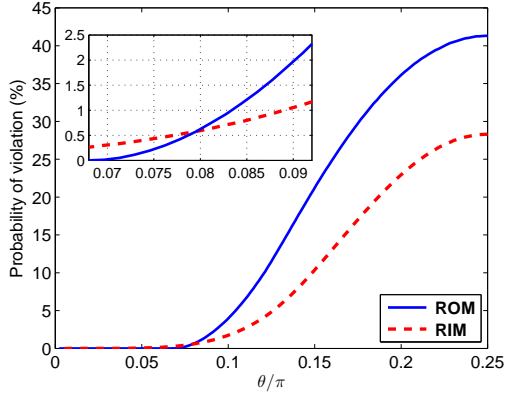}}
\caption{\label{Fig:ProbPureTwoQubit}
(Color online) Probability of finding the measurement statistics on a pure
two-qubit state $\ket{\Psi}=\cos\theta\ket{0}_1\ket{0}_2
+\sin\theta\ket{1}_1\ket{1}_2$ to violate the class of Bell-CHSH inequalities.
Here, the largest Schmidt coefficient $\cos\theta$ quantifies the entanglement.
The values $\theta=0$ and $\pi/4$ give, respectively, the pure product state
and the maximally entangled two-qubit state. Inset: a magnified plot of the
crossover region.}
\end{figure}

\subsection{Probability of violation vs entanglement}

For $n=2$, a pure quantum state can always be written as
$\ket{\Psi}=\cos\theta\ket{0}_1\ket{0}_2 +\sin\theta\ket{1}_1\ket{1}_2$ for
some local bases.  We numerically compute the probability of violation as a
function of entanglement, as shown in Fig.~\ref{Fig:ProbPureTwoQubit}.  Note
that, while the probability of violation is greater with ROM in most instances,
it also decreases more rapidly to zero with decreasing $\theta$.  In fact, at
$\theta\approx0.068\,\pi$, the probability of violation of $\ket{\Psi}$ is
already close to $10^{-5}\%$ with ROM whereas the chance is still about 0.26\%
with RIM (inset, Fig.~\ref{Fig:ProbPureTwoQubit}).


\begin{thebibliography}{99}
\bibitem{fn:AllMABK} All distinct inequalities in the MABK class can be
    generated by applying the $2^n$ different permutations ($s_k\to s_k$ or
    $s_k\to3-s_k$ for all $k=1,2,\ldots n$) to the coefficients
    $\beta(s_1,\ldots,s_n)$ of Eq.~\eqref{Eq:MABK:Coeff} defining the
    inequality.
\end{thebibliography}
\end{document}